\documentclass{article}  
\usepackage{amsmath}
\usepackage{todonotes}
\usepackage{bm}
\usepackage{cancel}
\usepackage{xcolor}   
\usepackage{arydshln}

\newcommand{\xxx}[1]{{\cancel{#1}}}
\newcommand{\BK}{\mathcal{I}}
\newcommand{\bfx}{{\bf{x}}}

\newcommand{\bfrho}{{\bm{\rho}}}
\usepackage{ulem}
\definecolor{grey}{RGB}{100,100,100}

\begin{document}                  



\title{Global Small-Angle Scattering Data Analysis of Inverted Hexagonal Phases}

\date{} 
\maketitle                        


\begin{center} 
\textsc{Moritz P.K. Frewein $^{a,b}$, Michael Rumetshofer $^c$, Georg Pabst $^{a,b}$}\\
\ \\
$^a$ University of Graz, Institute of Molecular Biosciences, Biophysics Division, NAWI Graz, 8010 Graz, Austria\\
$^b$ BioTechMed Graz, 8010 Graz, Austria\\
$^c$ Graz University of Technology, Institute of Theoretical Physics and Computational Physics,  NAWI Graz, 8010 Graz, Austria\\
georg.pabst@uni-graz.at
\end{center}

\section{Abstract}
We have developed a global analysis model for randomly oriented, fully hydrated inverted hexagonal (H$_\text{II}$) phases formed by many amphiphiles in aqueous solution, including membrane lipids. The model is based on a structure factor for hexagonally packed rods and a compositional model for the scattering length density (SLD) enabling also the analysis of  positionally weakly correlated H$_\text{II}$ phases. For optimization of the adjustable parameters we used Bayesian probability theory, which allows to retrieve parameter correlations in much more detail than standard analysis techniques, and thereby enables a realistic error analysis. The model was applied to different phosphatidylethanolamines including previously not reported H$_\text{II}$ data for diC14:0 and diC16:1 phosphatidylethanolamine. The extracted structural features include intrinsic lipid curvature, hydrocarbon chain length and area per lipid at the position of the neutral plane.

     
\section{Introduction}
Elastic small-angle scattering (SAS) techniques are unrivaled for providing detailed structural insight into aggregates formed by amphiphiles in aqueous solutions \cite{Glatter.2018}. In the field of membrane biophysics significant efforts have been devoted to the development of SAS analysis methods for the biologically most relevant fluid lamellar phases, including domain-forming lipid mixtures and asymmetric lipid bilayers \cite{Heberle.2017}.
In contrast, non-lamellar phases such as the inverted hexagonal (H$_\text{II}$) phase are less commonly found for membrane lipids under physiological conditions, but are of significant biotechnological interest, e.g. for gen transfection \cite{Koltover.1998} or drug delivery systems \cite{Yaghmur.2009}. H$_\text{II}$ phases are also highly amenable systems for deriving intrinsic lipid curvatures by small-angle X-ray scattering (SAXS) \cite{Leikin.1996,DiGregorio.2005,Kollmitzer.2013,Chen.2015}, which is the main focus of the present contribution. The intrinsic lipid curvature $C_0$ is given by the negative inverse of the curvature radius, $-1/R_0$, of an unstressed monolayer at the position of the neutral plane, which corresponds to the location where molecular bending and stretching modes are decoupled \cite{Kozlov.1991}. Major interest in obtaining reliable $C_0$-values originates from its contribution to the stored elastic energy strain in planar bilayers \cite{Marsh.2006}, transmembrane protein function \cite{Dan.1998,Frewein.2016} and overall membrane shape \cite{Frolov.2011}. 

Structural details of H$_\text{II}$ phases have been successfully derived using electron density map reconstruction based on Bragg peak scattering only \cite{Tate.1989,Turner.1992,Rand.1990,Harper.2001}. However, for highly swollen H$_\text{II}$ phases or at elevated temperatures the number of observed Bragg peaks may become insufficient for a reliable analysis. This may be particularly the case for mixtures of cone-shaped (H$_\text{II}$ phase-forming) and cylindrically-shaped (lamellar phase-forming) or inverted cone-shaped (spherical micelle-forming) lipids, which is the typical strategy for determining $C_0$ for non H$_\text{II}$ phase-forming lipids (see, e.g.~\cite{Kollmitzer.2013}). In this case global analysis techniques, which take into account both Bragg peaks and diffuse scattering become advantageous, as demonstrated previously also for lamellar phases \cite{Pabst.2000}. 

Global analysis techniques have been reported previously for H\textsubscript{I} phases, i.e., oil-in-water type hexagonal aggregates  \cite{Freiberger.2006,Sundblom.2009}.
The specific need for developing a dedicated model for H$_\text{II}$ phases comes from the observation that unoriented H$_\text{II}$ phases contain previously not reported additional diffuse scattering originating most likely from packing defects.
We have evaluated our global H$_\text{II}$ model for phosphatidylethanolamines with differing hydrocarbon chain composition and as a function of temperature using Bayesian probability theory to increase the robustness of analysis. This method  significantly increased the obtained information content compared to our previous analysis \cite{Kollmitzer.2013} and allowed us to derive details about the structure, e.g. the lipid headgroup area, hydrocarbon chain length or molecular shape to name but a few.


\section{Experimental methods}
\subsection{Sample preparation}
\noindent Dioleoyl phosphatidylethanolamine (DOPE, diC18:1 PE), palmitoyl oleoyl phosphatidylethanolamine (POPE, C16:0-18:1 PE), dimyristoyl phosphatidylethanolamine (DMPE, diC14:0 PE) and dipalmitoleoyl phosphatidylethanolamine (diC16:1 PE) were purchased in form of powder from Avanti Polar Lipids (Alabaster, AL). \textit{Cis}-9-tricosene was obtained from Sigma-Aldrich (Vienna, Austria). All lipids were used without any further purification. Note that dipalmitoleoyl phosphatidylethanolamine is deliberated abbreviated with diC16:1 PE in order to be not confused with dipalmitoyl phosphatidylethanolamine (diC16:0 PE). 

Fully hydrated H$_\text{II}$ phases were prepared using rapid solvent exchange (RSE) \cite{Buboltz.1999} as detailed previously \cite{Leber.2018}. In brief, stock solutions of lipids (10 mg/ml) and tricosene (5 mg/ml) were first prepared by dissolving both compounds in chloroform/methanol (9:1 vol/vol). Ultra pure water (18 M$\Omega/\text{cm}^2$) was filled into test tubes and equilibrated at (60-$70)^\circ$C using an incubator. Lipid and tricosene stock solutions were added to the test tubes containing preheated water (organic solvent/water ratio = 2.55) and an then quickly mounted onto the RSE apparatus, described in \cite{Rieder.2015}. Organic solvent quickly evaporated using the following settings:  temperature: 65 $^\circ$C; vortex speed: 600 rpm; argon-flow: 60 ml/min and a final vacuum of pressure: (400-500) mbar. The full procedure was performed for 5 minutes, yielding a lipid pellet at the bottom of the test tube in excess of water. All samples contained 12 wt.\% tricosene. Tricosene inserts preferentially into the interstical space between the rods in H$_\text{II}$ phases 
effectively reducing packing frustration as verified previously \cite{Alley.2008,Kollmitzer.2013}. Unstressed H$_\text{II}$ phases are required for $C_0$ determination~\cite{Kozlov.1991}.


\subsection{Scattering experiments}
\noindent Small angle X-ray scattering (SAXS) experiments were performed on a SAXSpace compact camera (Anton Paar, Graz, Austria) equipped with an Eiger R 1 M detector system (Dectris, Baden-Daettwil, Switzerland) and a 30 W-Genix 3D microfocus X-ray generator
(Xenocs, Sassenage, France), supplying Cu-K$\alpha$ ($\lambda =$ 1.54 \AA) radiation with a circular spot size of the beam of $\sim$300 $\mu$m on the detector. Samples were taken-up in paste cells (Anton Paar) and equilibrated at each measured temperature for 10 minutes using a Peltier controlled sample stage (TC 150, Anton Paar). The total exposure time was 32 minutes (4 frames of 8 min), setting the sample-to-detector distance to 308 mm. Data reduction, including sectorial data integration and corrections for sample transmission and background scattering, was performed using the program SAXSanalyis (Anton Paar).


\section{Model}
\subsection{General aspects for H$_\text{II}$ phases}
We initially tested the applicability of a previously reported model-free approach \cite{Freiberger.2006}. However, although perfect fits to the experimental data were obtained, the corresponding pair distance distribution functions contained significantly negative values upon approaching the maximum particle size ($D_{max}$), which is not physically relevant (O. Glatter, personal communication). This encouraged us to proceed with data modeling.

\begin{figure}\label{fig:lat_un}
\caption{Scheme of the H$_\text{II}$ phase model. The hexagonal lattice (left side) is defined by its lattice parameter $a$ and the number of rings (lattice order) $n$. Its unit cell, shown in the center, is a regular hexagonal prism of length $L$ and consists of a cylindrical water core, surrounded by lipids with their heads pointing toward the central water channel and a filler molecule occupying the interstices. We denote the axis of rotation by $z$. The unit cell is separated in areas of different SLD which depend on the molecular composition (see also Fig.~\ref{fig:lipunit}).  $R_0$, denotes the position of the neutral plane at the center of the lipid backbone.}
\includegraphics[width=\textwidth]{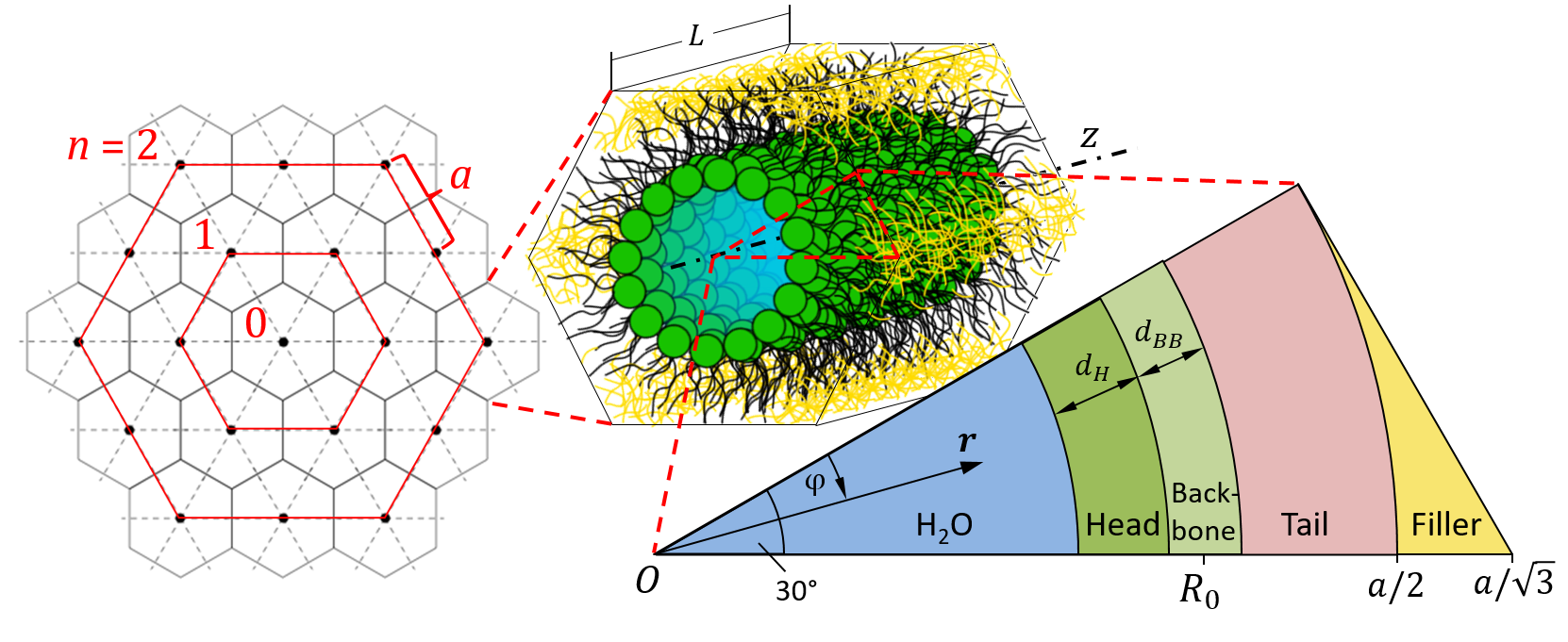}
\end{figure}

To do so, we considered a bundle of hexagonal prisms consisting of a water core coated by lipids (Fig.~\ref{fig:lat_un}). Its scattering intensity is characterized by the form factor of a single prism $F(\mathbf{q})$ and the structure factor of the whole bundle $S(\mathbf{q})$, where $\mathbf{q}$ is the scattering vector. Assuming that the prisms are long as compared to their diameters allows us to decouple form and structure factor \cite{Freiberger.2006}:
\begin{equation}\label{eq:form_struc3}
I (\mathbf{q})\propto |F(\mathbf{q})|^2 S(\mathbf{q}),
\end{equation}

The structure factor of a two-dimensional lattice of infinitely long hexagonal prisms, averaged over all in-plane vectors is given by \cite{Oster.1952,Marchal.2003,Freiberger.2006}:
\begin{equation}\label{eq:S}
S(q, \theta| n,\Delta) =1 + \frac{1}{N_{\text{hex}}(n)}e^{-q^2 \Delta}\sum_{j\neq k}^{N_\text{hex}(n)}  J_0(q|\mathbf{R}_j - \mathbf{R}_k|\sin\theta),
\end{equation}
where $\theta$ is the angle between scattering vector and the axis ($z$) normal to the hexagons, 
$N_{hex} = 1+3n(n+1)$ is the total number of unit cells for $n$ rings (Fig.~\ref{fig:lat_un}), and $J_0$ is the zero-order Bessel function of first kind. The $e^{-q^2 \Delta}$-term is well-known as Debye-Waller factor, where $\Delta$ is the lateral mean square displacement of the rotation axes of the unit cells around their mean positions  $\mathbf{R}_j$. 
For the sake of readability we give the parameter dependencies after the vertical line in each equation, i.e. for Eq.~(\ref{eq:S}) $n$ and $\Delta$ in.
Analogously to \cite{Freiberger.2006} we also considered a polydispersity of $N_\text{hex}$, which yields a smooth structure factor. However, since this affects only low $q$-vectors and not significantly the final quality of the result, this was omitted in order to reduce overall computation times.
An alternative structure factor, based on the positioning of peaks with flexible shapes on hexagonal lattice points, has been reported previously \cite{Forster.2005,Sundblom.2009}. However, its application involves a significantly higher number of adjustable parameters, which lead us to disregard this option.

The form factor for a hexagonal prism of length $L$ is given by \cite{Freiberger.2006}
\begin{eqnarray} \label{eq:ff}
F(q, \theta|\bfrho) & = & f(q,\theta) \int \rho(r,\varphi) J_0(q r \sin\theta) rdrd\varphi \nonumber \\ 
    & = & f(q,\theta) \left[ F_{\text{lipid}}(q, \theta|\bfrho) + F_{\text{inter}}(q, \theta|\bfrho) \right]
\end{eqnarray}
with $f(q, \theta) = 4\pi\sin(\frac{L}{2}q\cos\theta)/(q\cos\theta)$, where $\rho(r,\varphi) $ is the in-plane SLD.
Here, $\bfrho$ denotes all parameters describing the SLD $\rho(r,\varphi)$. For evaluation, which is performed due to symmetry over $1/12$ of the area of the hexagon, the form factor is split into a core-shell cylindrical part $F_\text{lipid}$, which accounts for the phospholipid only, and $F_\text{inter}$, which accounts for the interstitial space, often taken up by pure hydrophobic filler molecules (here: tricosene). We also found that the length of the cylinders does not affect $F$ significantly for $L \geq 2500$ \AA\/ for cylinder radii between 35 and 45 \AA, as occurring in the present paper. For shortening computational times, we therefore fixed $L = 2500$ \AA\/ for all our further calculations.

The core-shell cylindrical part can be evaluated analytically \cite{Szekely.2010}
\begin{eqnarray}
F_\text{lipid}(q,\theta|\bfrho) & = &  \int_0^{a/2} r\Delta\rho(r) J_0(q r \sin\theta) dr =\nonumber\\ 
& = &  \frac{1}{q \sin{\theta}}\Bigg[ \Delta\rho_M r_M J_1(q r_M \sin\theta) + \nonumber \\
&  & + \sum_{k=1}^{M-1} (\Delta\rho_k - \Delta\rho_{k+1})r_k J_1(qr_k \sin\theta)\Bigg], 
\end{eqnarray}
where $M$ is the total number of shells, $r_k$ are the shell radii, $\Delta\rho$ is the SLD relative to water ($\Delta\rho = \rho - \rho_{W}$; $\rho_{W} = 0.33$ \AA$^{-3}$ in case of X-rays), and $J_1$ is the first-order Bessel function of the first kind. 

Molecular fluctuations cause a smearing of the sharp boundaries between the individual slabs. Analogously to \cite{Franks.1982}, these were taken into account by translating all shell boundaries $\{r_k\}$ by the distance $x$, whose value was assumed to be distributed by a Gaussian $\mathcal{N}(x|\mu,\sigma_{\text{fluc}}^2)$ of mean $\mu$ and variance $\sigma_\text{fluc}^2$.
%
%
\begin{equation}
F_\text{lipid,~fluc}(q,\theta|\bfrho,\sigma_\text{fluc}) = \int dx \; \mathcal{N}(x|\mu=0,\sigma_\text{fluc}^2)~F_\text{lipid}(q,\theta|\bfrho'(x) )
\end{equation}
%
Here, $\mu=0$ and $\bfrho'(x)$ denotes the SLD including the radial shift $x$.

The form factor of the interstices 
\begin{equation}
\label{eq:formfac_inter}
F_\text{inter}(q,\theta|\bfrho) = \frac{6}{\pi} \int_0^{\pi/6} d\varphi \int_{a/2}^{\frac{a}{2\cos(\varphi)}} r\Delta\rho_\text{inter} J_0(qr \sin\theta) dr 
\end{equation}
needs to be evaluated numerically, but remains constant for a given lattice constant $a$ and SLD $\Delta\rho_\text{inter}$. However, since $a$ can be  determined accurately from Bragg peak positions, $F_\text{inter}$ needs to be calculated only once for each scattering pattern. 

H$_\text{II}$ phases are well-known to change their phase from '$+$' to '$-$' between the (1,0) and (1,1) reflections \cite{Turner.1992}, which brings about a minimum in $|F_\text{lipid}|^2$ between the two peaks.

\begin{figure}\label{fig:inten+formf}
\caption{Overlay of the scattering pattern of DOPE (circles) and $|F_\text{lipid}(q)|^2$ (solid line). The phase change between the (1,0) and (1,1) reflections for the H$_\text{II}$ phase leads to a minimum in the absolute square of the form factor, which is absent in the experimental data.}
\includegraphics[width=\textwidth]{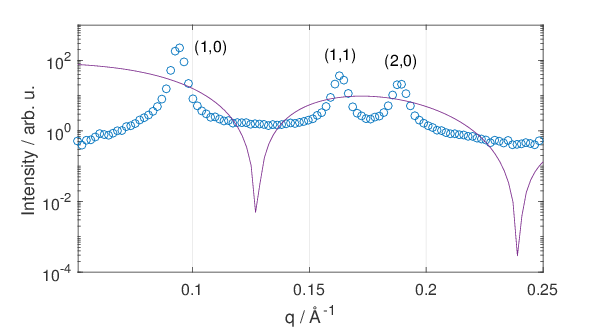}
\end{figure}

All our present experimental data, as well as those previously reported~\cite{Kollmitzer.2013,Leber.2018}, exhibited significant diffuse scattering between these two peaks (Fig.~\ref{fig:inten+formf}). That is, experimental data from unoriented H$_\text{II}$ phases show no form factor minimum in this $q$-range. The additional scattering may also explain the failure of the model-free analysis approach discussed above and possibly arises from packing defects between hexagonal bundles, e.g. at grain boundaries. However, surface-aligned H$_\text{II}$ phases do not exhibit such scattering contributions \cite{Ding.2004}, disfavoring such a scenario. Hence, this appears to be only a property of unoriented H$_\text{II}$ phases, fully immersed in aqueous solution. We speculate that the outermost boundary of H$_\text{II}$ structures may try to avoid contact of the hydrocarbon with water by forming a lamellar layer, i.e., in some ways similar to hexosomes \cite{Yaghmur.2009}. Indeed, we were able to account for the additional diffuse scattering adding a form factor of a laterally uniform, infinitely extended, planar bilayer
\begin{equation}\label{eq:bilff2}
F_\text{BL}(q|\bfrho_\text{lam}) = 4\pi^2 \int \Delta \rho_\text{lam}(z) e^{iqz}~dz
\end{equation}
to the total scattered intensity, where $z$ is the coordinate normal to the lamellar phase and $\bfrho_\text{lam}$ are the parameters describing the SLD of the lamellar phase. We cannot exclude that the additional diffuse scattering originates from unilamellar vesicles or other kinetically-trapped aggregates formed during sample preparation.

Considering orientational averaging we finally obtained for the total scattered intensity
\begin{align}
\label{eq:total_intensity}
 I^\text{mod}(q| n,\Delta, \sigma_\text{fluc},& \bfrho,c_\text{lam},\bfrho_\text{lam}) \propto \int_0^\pi |F(q,\theta|\bfrho,\sigma_\text{fluc})|^2 S(q,\theta|n,\Delta) \sin\theta d\theta~+ \nonumber \\ 
& +~2c_\text{lam}~F_\text{BL}(q|\bfrho_\text{lam}) \int_0^\pi F(q,\theta|\bfrho,\sigma_\text{fluc}) s(q,\theta|n,\Delta) \sin\theta d\theta~+ \nonumber \\
& +~c_\text{lam}^2~|F_\text{BL}(q|\bfrho_\text{lam})|^2,
\end{align}
where $c_\text{lam}$ denotes the fraction of the lamellar phase. The structure factor
\begin{equation}
s(q, \theta|n,\Delta) =\frac{1}{\sqrt{N_\text{hex}(n)}} e^{-q^2 \Delta/2}\sum_{j}^{N_\text{hex}(n)}  J_0(q |\mathbf{R}_j| \sin\theta)
\end{equation}
was derived analogously to the H$_\text{II}$ structure factor (Eq.~\ref{eq:S}) and the form factor is $F(q,\theta|\bfrho,\sigma_\text{fluc})=f(q,\theta) \left[ F_\text{lipid,~fluc}(q,\theta|\bfrho,\sigma_\text{fluc}) + F_\text{inter}(q,\theta|\bfrho)\right]$, see Eq.~\ref{eq:ff}.

\subsection{Composition-specific modeling}

\begin{figure}\label{fig:lipunit}
\caption{Composition-specific SLD modeling of phosphatidylethanolamines. a) The unit cell of a single lipid has the shape of a cylinder sector of radius $a/2$ . b) Parsing of DOPE into head (H), backbone (BB) and hydrocarbon chain (HC) and chemical structure of the tricosene. c) Scheme of a corresponding electron density profile (see also Fig.~\ref{fig:lat_un}).}
\includegraphics[width=\textwidth]{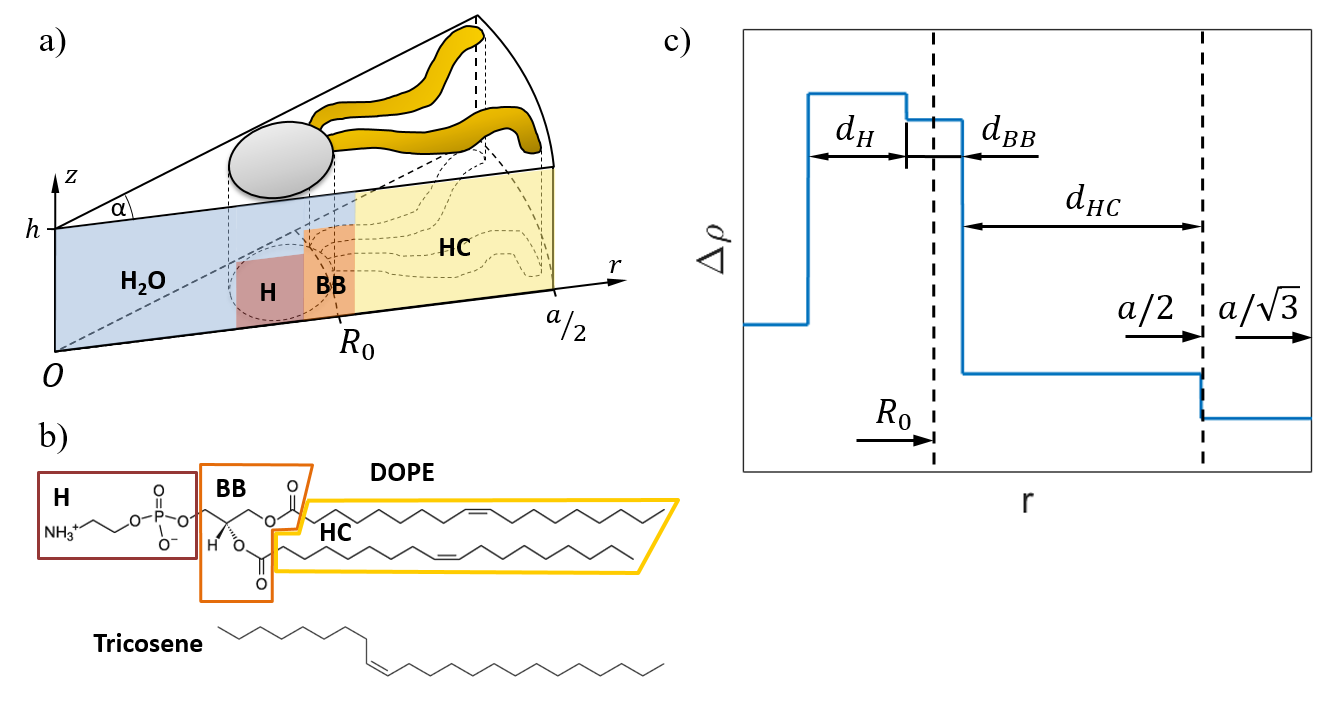}
\end{figure}

In this section we develop a model for the SLDs described by the parameters $\bfrho$ and $\bfrho_\text{lam}$. For increased structural fidelity we considered the minimum amount of parameters. We also constrained the SLDs by the specific molecular composition. Assuming that tricosene partitions exclusively into the interstitial space, the PE structure was parsed into three cylindrical shells of a wedge-shaped lipid unit cell of opening angle $\alpha$ and height $h$ (Fig.~\ref{fig:lipunit}): (i) the headgroup (H), consisting of phosphate and ethanolamine groups, (ii) the glycerol backbone (BB), given by the carbonyl and glycerol groups, and (iii) the tails (HC) consisting of all methyl, methine and methylene groups. The outer radius of the wedge $a/2$ is evaluated in advance from the Bragg peak positions $q_{kl} = \frac{4\pi}{\sqrt{3}a}(k^2+2kl+l^2)$, where $k$ and $l$ are the Miller indices. The position of the neutral plane $R_0$ was assumed to be in the center of the BB shell. This was motivated by bending/compression experiments, which obtained estimates for the location of neutral plane within the lipid backbone regime~\cite{Kozlov.1991,Leikin.1996}. In our model, the entire PE structure is described by the intrinsic curvature $C_0=-1/R_0$, the width of the headgroup $d_\text{H}$ and the backbone $d_\text{BB}$. Further structural parameters of interest, as the width of the hydrocarbon chain 
\begin{equation}
	d_\text{HC}= a/2-d_\text{BB}/2-R_0 \;,
\end{equation}
the lipid head-to-headgroup length
\begin{equation}
	d_\text{HH}= 2(d_\text{H}+d_\text{BB}+d_\text{HC})
\end{equation}
and the radius of the water core
\begin{equation}
	R_\text{W} = (a - d_\text{HH})/2 = R_0-d_\text{BB}/2-d_\text{H}\;,
\end{equation}
follow from these three parameters.

In the case of X-ray scattering the SLDs (electron densities) of each shell are given by $\rho_k = n^e_k/V_k$ with $k\in\ \{ \text{H, BB, HC}$\}, where $n^e_k$ is the number of electrons of a given quasi-molecular lipid fragment, $V_\text{H} = 110$ \AA$^3$, $V_\text{BB} = 135$ \AA$^3$ \cite{Kucerka.2015}, and $V_\text{HC}=V_\text{lipid}-V_\text{BB}-V_\text{H}$. Further, we estimated $\Delta\rho_\text{inter}$ (Eq.~(\ref{eq:formfac_inter})) of tricosene by molecular averaging over the fractional volumes of $v_\text{CH}$, $v_{\text{CH}_2}$ and $v_{\text{CH}_3}$ \cite{Kucerka.2015} (see supporting Tab. S1). 
%
In our model the electron density is sufficiently described by the parameters $C_0$, $d_{\text{H}}$, $d_{\text{BB}}$ and $V_{\text{lipid}}$, hence $\bfrho = \{ C_0, d_{\text{H}}, d_{\text{BB}}, V_{\text{lipid}}\}$. All other parameters can be deduced from these by using the lipid contribution to volume of the $k$'th shell,
\begin{equation}
\label{eq:shell_volume}
V_k = \frac{\hat{A}~(r_{k+1}^2-r_k^2)}{2} - \tilde{n}_\text{W}^k V_\text{W} \;,
\end{equation}
where $\tilde{n}_\text{W}^k$ is the number of water molecules within each shell and $V_\text{W} = 30$ \AA$^3$ the molecular volume of water. $\hat{A} = \alpha h$ is the mantle area of a sector of unitary radius and can be obtained using Eq.~(\ref{eq:shell_volume}) with $k=\text{HC}$ and $\tilde{n}_\text{W}^\text{HC} = 0$,
\begin{equation}
\hat{A}=\frac{2V_\text{HC}}{\frac{a^2}{4}-(R_0 + \frac{d_\text{BB}}{2})^2}.
\end{equation}
Hence, Eq.~(\ref{eq:shell_volume}) also defines $\tilde{n}_\text{W}^\text{H}$ and $\tilde{n}_\text{W}^\text{BB}$. 
%
%

Using our parametrization it is straight forward to derive the area per lipid at any position within the molecule. For example, the area per lipid at the neutral plane calculates as
\begin{equation}
	A_0 = \hat{A}R_0.
\end{equation}
Further, following~\cite{Israelachvili.2011}, the molecular shape parameter is given by
\begin{equation}
	\tilde{S} = \frac{V_\text{HC}}{\hat{A}(R_0+\frac{d_\text{BB}}{2}) d_\text{HC}},
\end{equation}
where $\tilde{S}=1$ represents cylindrical -- lamellar phase forming -- molecules, and $\tilde{S}<1$ or $\tilde{S}>1$ typify molecules inducing negative or positive monolayer curvatures, respectively. In particular, $\tilde{S} > 1$ for amphiphiles forming aggregates with negative curvature, like the H$_\text{II}$-structure.

The form factor of the additional lamellar phase was calculated by integrating Eq.~(\ref{eq:bilff2}), using a simple SLD model consisting of head and tail slabs~ \cite{Szekely.2010}:
\begin{eqnarray}
F_\text{BL}(q) &= &\frac{4\pi^2}{iq}\Big\{\Delta\rho_\text{H,lam}\left[\operatorname{e}^{\mathrm{i} q d_\text{H,lam}}- 1 \right] + \nonumber \\
& + & \Delta\rho_\text{H,lam} \left[\operatorname{e}^{\mathrm{i} 2 q (d_\text{H,lam}+d_\text{HC,lam})} - \operatorname{e}^{\mathrm{i} q(d_\text{H,lam}+2d_\text{HC,lam}) }\right] + \nonumber \\
& + & \Delta\rho_\text{HC,lam}\left[\operatorname{e}^{\mathrm{i} q (d_\text{H,lam}+2d_\text{HC,lam})} - \operatorname{e}^{\mathrm{i} q d_\text{H,lam}}\right]
\Big\},
\end{eqnarray}
where $\Delta \rho_\text{H,lam}$ and $\Delta \rho_\text{HC,lam}$ are the headgroup and hydrocarbon SLDs relative to water, respectively. These were derived as detailed above by counting the number of electrons in each slab and dividing by the corresponding volumes $V_\text{H,lam}$ or $V_\text{HC,lam}$. Assuming that $V_\text{HC,lam} = V_\text{HC}$ and $V_\text{H,lam} = V_\text{H} + V_\text{BB}$, the hydrocarbon slab thickness results from 
\begin{equation}
d_\text{HC,lam} = \frac{V_\text{HC}}{A_\text{L}} 
\end{equation}
and the headgroup thickness from
\begin{equation}
d_\text{H,lam} = \frac{V_\text{H,lam}+n_\text{W,lam} V_\text{W}}{A_\text{L}}.
\end{equation}
Hence, the area per lipid $A_\text{L}$ and the number of headgroup-associated water molecules $n_\text{W,\,lam}$ are the only parameters for the lamellar phase, $\bfrho_{\text{lam}}=\{ A_\text{L},n_\text{W,\,lam}\}$. 

\section{Parameter estimation using Bayesian probability theory}

The final model for scattered intensities of unoriented fully hydrated H$_\text{II}$ is given by
\begin{equation}
I^{\text{sim}}(q|\bfx )  = \Gamma I^{\text{mod}}(q| n,\Delta, \sigma_\text{fluc} ,\bfrho,c_\text{lam},\bfrho_\text{lam}) + I_\text{inc},
\end{equation}
where $I^{\text{mod}}$ is given by Eq.~(\ref{eq:total_intensity}), $\Gamma$ is an instrumental scaling constant and $I_\text{inc}$ accounts for incoherent scattering. In total, we have 12 model parameters, denoted by $\bfx$
, which are listed in Tab.~\ref{tab:fitpar}.

\begin{table}\label{tab:fitpar}
\caption{Overview of the model parameters for fully hydrated unoriented H$_\text{II}$ phases.}
\begin{tabular}{lc|l}      
Occurrence & $\bfx$ & Meaning  \\
\hline
Structure factor& $\Delta$ & mean square displacement of the lattice points\\
 & $n$ & number of hexagonal shells (domain size)\\ \hdashline
H$_\text{II}$ form factor & $\sigma_\text{fluc}$ & fluctuation constant of lipid unit cell\\
 & $C_0$ & intrinsic curvature \\
 & $d_\text{H}$ & width of the lipid headgroup\\
 & $d_\text{BB}$ & width of the lipid backbone\\
 & $V_\text{lipid}$ & lipid volume \\ \hdashline
Lamellar form factor & $c_\text{lam}$ & lamellar form factor scaling constant \\ 
 & $A_\text{L}$ & area per lipid of the lamellar phase \\
 & $n_\text{w,lam}$ & number of water molecules in the headgroup slab of the lamellar phase\\ \hdashline
Signal scaling & $\Gamma$ & instrumental scaling constant\\
 & $I_\text{inc}$ & incoherent background\\ 
\hline
\end{tabular}
\end{table}

There are various ways of estimating the parameters $\bfx$. For a given data set $\mathbf{I}$ with the standard deviations $\bm{\sigma}$ in the presence of a well-defined global minimum the method of least squares yields fitting parameters by minimizing a cost function $\chi^2(\bfx|\mathbf{I},\bm{\sigma})$. However, such an approach led for our present data to a significant variation of results between consecutive optimization runs, indicating a cost function landscape with a weakly-defined global minimum. Thus, besides unreliable $\bfx$ values, also error estimates and potential correlations between the parameters remained undetermined.

To achieve higher confidence in our results we decided to use Bayesian probability theory; for a detailed introduction, see \cite{Jaynes.2003,Gregory.2005,Sivia.2012,Linden.2014}. In brief, we were interested in deriving the probability $p(\bfx|\mathbf{I},\bm{{\sigma}},\BK)$, meaning the probability the parameters $\bfx$ given the set of experimental data $\mathbf{I}$ with standard deviations $\bm{\sigma}$ and additional information $\BK$, which might be present, such as e.g. the finite width of a lipid molecule. In the framework of Bayesian probability theory Bayes' theorem shows how to calculate this quantity, also called the posterior,
\begin{align}
\overbrace{p(\bfx|\mathbf{I},\bm{{\sigma}},\BK)}^{\text{posterior}} \propto \overbrace{p(\mathbf{I}|\bfx,\bm{{\sigma}},\BK)}^{\text{likelihood}} \overbrace{p(\bfx|\xxx{\bm{{\sigma}}},\BK)}^{\text{prior}} \;.
\end{align}
Bayes' theorem constitutes the rule for learning from experimental data. The prior probability $p(\bfx|\BK)$ represents the prior knowledge about the unknown quantities $\bfx$. We crossed out $\bm{{\sigma}}$ in the prior, since the prior does not depend on the standard deviations of the data. The likelihood $p(\mathbf{I}|\bfx,\bm{{\sigma}},\BK)$, representing the probability for the data $\mathbf{I}$ given $\bfx$ and $\bm{{\sigma}}$, includes all information about the measurement itself.
The prior probabilities $p(\bfx|\BK)$ were assumed to be uniformly distributed between lower $\bfx_{\text{min}}$ and upper $\bfx_{\text{max}}$ constraints for all parameters. Therefore,
\begin{equation}\label{eq:prior}
p(\bfx|\BK) = \prod_i\frac{\Theta(x_i-x_{i,\text{min}}) - \Theta(x_i-x_{i,\text{max}})}{x_{i,\text{max}}-x_{i,\text{min}}}\;,
\end{equation}
where $\Theta(x)$ is the Heaviside step function. For each parameter $x_i$, $x_{i,\text{min}}$ and $x_{i,\text{max}}$ denote physically meaningful boundaries. In particular, we constrained the parameters $d_\text{H}$ and $d_\text{BB}$ by the conditions 
\begin{equation}\label{eq:nw}
\tilde{n}_W^\text{H}\geq 0 \text{~~and~~} \tilde{n}_W^\text{BB}\geq 0. 
\end{equation}
This means that the volumes of the head and backbone shell (Eq.~(\ref{eq:shell_volume})) have to be large enough to accommodate the respective molecular group.
%

We consider the likelihood $p(\mathbf{I}|\bfx,\bm{{\sigma}},\BK)$. Since we did not trust \textit{per se} the experimentally derived error estimates $\bm{{\sigma}}$ for the scattered intensities we assumed that their real values $\bm{\tilde{\sigma}}$ are connected to $\bm{{\sigma}}$ by a scaling factor $\eta$. Using the marginalization rule of Bayesian probability theory we obtain
\begin{align} \label{eq:scaleeta}
p(\mathbf{I}|\bfx,\bm{\sigma},\BK) &= \int d\eta d\bm{\tilde\sigma}~p(\mathbf{I}|\bfx,\xxx{\bm{\sigma}},\bm{\tilde{\sigma}},\xxx{\eta},\BK)p(\bm{\tilde{\sigma}},\eta|\xxx{\bfx},\bm{\sigma},\BK) \notag \\
&= \int d\eta d\bm{\tilde\sigma}~p(\mathbf{I}|\bfx,\bm{\tilde{\sigma}},\BK)\underbrace{p(\bm{\tilde{\sigma}}|\eta,\bm{\sigma},\BK)}_{\prod_i\delta(\tilde\sigma_i-\eta\sigma_i)}\underbrace{p(\eta|\xxx{\bm{\sigma}},\BK)}_{\propto 1/\eta} \notag \\
&\propto \int d\eta ~ \underbrace{p(\mathbf{I}|\bfx,\bm{{\tilde\sigma}}=\eta\bm{{\sigma}},\BK)}_{\mathcal{N}(\mathbf{I}|\bfx,\eta\bm{{\sigma}})}\frac{1}{\eta}\;,
\end{align}
where $d\bm{\tilde\sigma}$ is short hand for $\prod_i d{\tilde\sigma}_i$.
Here, we have specifically made use of the so-called Jeffreys prior $p(\eta)\propto 1/\eta$  \cite{Jeffreys.1946}, where $\eta$ is a scaling invariant, meaning that we have \textit{a priory} no idea about the order of magnitude of $\eta$. 
This scaling connects the likelihood (Eq.~(\ref{eq:scaleeta})) to the multivariate Gaussian
\begin{equation}
\mathcal{N}(\mathbf{I}|\bfx,\eta\bm{\sigma}) = \prod_{i=1}^{N_q} \frac{1}{\eta\sigma_i\sqrt{2\pi}}\exp\left[-\frac{1}{2\eta^2\sigma_i^2}(I^{\text{sim}}(q_i|\bfx )-I_i^{\text{obs}})^2\right]
\end{equation}
where $\eta$ has to be integrated out, respecting Jeffreys prior and $I_i^{\text{obs}}$ denotes the observed intensity at $q_i$.
Here, $N_q$ is the number of data points for a given scattering pattern.

For illustration, consider an arbitrary function $\mathcal{O}(\bfx)$ with the parameters $\bfx$. The expectation value of $\mathcal{O}(\bfx)$ is then calculated by evaluating the integral
\begin{align}
\left< \mathcal{O}(\bfx) \right> = \int d\bfx d\eta ~ \mathcal{O}(\bfx) ~ p(\bfx,\eta|\mathbf{I},\bm{\sigma},\BK) \;
\label{eq:expectationMCMC}
\end{align}
where 
\begin{align}
p(\bfx,\eta|\mathbf{I},\bm{\sigma},\BK) = \frac{1}{Z}\mathcal{N}(\mathbf{I}|\bfx,\eta\bm{\sigma})\frac{1}{\eta} p(\bfx|\BK)\;
\end{align}
with the normalization constant $Z$.
For example, using $\mathcal{O}(\bfx) = x_i$ produces the expectation value $\left< x_i\right>$ for parameter $x_i$.
%

A suitable technique for performing these integrals and sampling from the probability distribution $p(\bfx,\eta|\mathbf{I},\bm{\sigma},\BK)$ is Markov Chain Monte Carlo (MCMC), which is based on constructing a so called Markov chain with the desired distribution of $\bfx$ in equilibrium. We used the Metropolis Hastings algorithm for generating the Markov chain $\{ \bfx^k, \eta^k \}$.
Starting with a parameter set $\bfx^{k=1}$ and $\eta^{k=1}$, every new parameter set $k+1$ can be proposed by varying parameters in the old parameter set $k$. The new parameter set $k+1$ is accepted with the probability
\begin{equation}
P_\text{acc} = \text{min}\left\{ 1, \frac{p(\bfx^{k+1},\eta^{k+1}|\mathbf{I},\bm{\sigma},\BK)}{p(\bfx^{k},\eta^{k}|\mathbf{I},\bm{\sigma},\BK)} \right\} \;.
\end{equation}

It occurred that the first $10-20$~\% of a Markov chain have to be discarded to ensure that the rest of the Markov chain is independent of the initial state $\bfx^{k=1}$ and $\eta^{k=1}$, i.e. the Markov chain is equilibrated to the desired distribution. 
In addition, the states in the Markov chain have to be uncorrelated, which can be ensured by taking only every $N_{\text{run}}$$^\text{th}$ state of the Markov chain.  $N_{\text{run}}$ can be controlled by evaluating the autocorrelation function or using techniques like binning and jackknife. 
Finally, the observable can be estimated by
\begin{equation}
\label{eq:mean}
 \mathcal{O} := \left< \mathcal{O}(\bfx) \right> \approx \frac{1}{N_{\text{Markov}}} \sum_{k=1}^{N_{\text{Markov}}} \mathcal{O}(\bfx^{k}) \;,
\end{equation}
i.e. taking the mean value of $N_{\text{Markov}}$ uncorrelated Markov Chain elements.
The confidence intervals can be estimated from
\begin{equation}
\Delta \mathcal{O}:= \frac{\sigma_{\mathcal{O}}}{\sqrt{N_{\text{Markov}}}}\;
\end{equation}
The variance 
\begin{equation}
\sigma_{\mathcal{O}}^{2}= \left<\mathcal{O}(\theta)^{2} \right> - \left<\mathcal{O}(\theta) \right>^{2} 
\label{eq:error_estimate}
\end{equation} 
can in turn be estimated from the Markov chain. Alternatively, the uncertainty can be determined from independent MCMC runs.

Since the Markov chain $\{ \bfx^k, \eta^k \}$ is a representative sample drawn from $p(\bfx,\eta|\mathbf{I},\bm{\sigma},\BK)$ it can be used to plot the probability distribution, e.g. the marginal probability distribution $p(x_i,x_j|\mathbf{I},\bm{\sigma},\BK)$ for the parameter $i$ and $j$ by plotting the two dimensional histogram of the samples $\{x^k_i\}$ and $\{x^k_j\}$. This allows to unravel correlations between the parameter $i$ and $j$, i.e. the analysis of mutual parameter dependencies that could lead to ambiguous results using the least squares method.
Additionally, the cost function
\begin{equation}
	\chi^2 =  \frac{1}{N_q}\sum_{i=1}^{N_q} \frac{\left[ I(q_i|\bfx)^\text{sim}-I_j^\text{obs} \right]^2}{\tilde{\sigma_i}^2}
\end{equation}
is saved for every run.

\section{Results and discussion}
\subsection{Tests of the analysis}
We first explored our model and the Bayesian analysis on the well-studied H$_\text{II}$ structure of DOPE \cite{Turner.1992,Harper.2001,Tate.1989,Kollmitzer.2013}.
We emphasize that the choice of our model restricts the algorithm to a certain functional space for describing the scattering pattern.
This constraint can lead to some systematic deviations from the experimental data and Bayesian model comparison can be used for choosing the appropriate model.

\begin{figure}\label{fig:fitdet}
\caption{Expectation value and error bands of the intensity of fully hydrated DOPE at 35 $^\circ$C (a) including the involved structure (b) and form factors (c, blue: hexagonal FF, green: lamellar FF).}
\includegraphics[width=\textwidth]{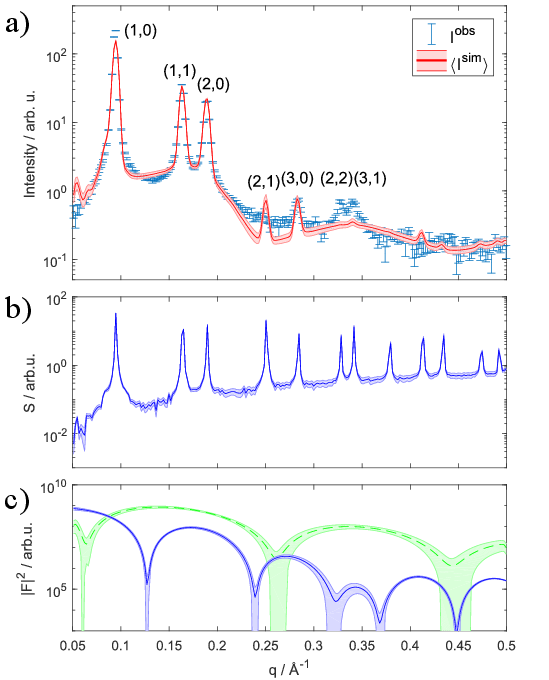}
\end{figure}

Clearly, our model is able to account well for most features of the scattering pattern up to $q \sim 0.5$ \AA$^{-1}$ (Fig. \ref{fig:fitdet}). In particular, the bilayer form factor compensates well for the form factor minimum between the (1,0) and the (1,1)-peak, but adds also some diffuse scattering at higher $q$-values. The small peak observed in the calculated intensity at very low $q$ is an artifact resulting from the structure factor. This could be removed by averaging over a distribution of domains~\cite{Freiberger.2006}, but does not affect the overall structural results and has therefore been omitted to reduce computational cost (see also above).
Further, the proximity of a form factor minimum to the (2,1)-peak of the H$_\text{II}$ phase nearly causes an extinction in the scattering data. Note that tricosene-free DOPE samples exhibit a clear (2,1) reflection (Fig. S4c). However, because of strain-induced distortions of the hexagonal prisms such samples cannot be analysed with the present model.
The maximum aposterior (MAP) solution still shows a slightly more pronounced (2,1)-peak, since a perfect fit in this $q$-range would lead to significant deviations between model and experimental data close to the (2,0)-peak, which due to its smaller errors have a higher significance in contributing to our overall goodness of the MAP solution. Additionally, our MAP solution underestimates the contributions of the (2,2) and (3,1) peaks due to the proximity of the cylinder form factor to two minima. 
To account for this we tested more complex SLD models, by considering either a separate slab for the methyl terminus of the hydrocarbon chain, or a linear decrease of the electron density in the hydrocarbon regime. However, this did not lead to a significant improvement of the agreement between model and experimental data in this $q$-range. In order to avoid overfitting we therefore remained with the SLD model as described in section 3.
Table S2 lists the corresponding expectation values $\langle\bfx\rangle$ and variances $\sigma_{\bfx}$. In order to check for reproducibilty, we prepared a fresh DOPE sample. Results listed in Tab.~S2 show that all structural lipid parameters are identical within experimental uncertainty (see also Fig.~S4b). Differences in lattice parameters, such as $a$ and $\Delta$ relate to slight variations of tricosene content.

One of the benefits of the Bayesian analysis compared to the least squares method is the possibility to reveal correlations between adjustable parameters, just by looking at the 2D marginal probability density distributions, see e.g. Fig. \ref{fig:corr}a. Marginal distributions of all other parameters are shown in the supplementary Figs.~S1-S3. Most parameter pairs show no correlations and exhibit probability distributions with Gaussian-like behavior, including $\sigma_{fluc}$, $V_\text{lipid}$, and $\Gamma$. Significant correlations can be seen for the parameters $d_\text{H}$ and $d_\text{BB}$ with $C_0$, as well as between $n_\text{w,lam}$ and $A_\text{L}$.
A strong correlation between two parameter suggests the possibility to simplify the model. However, this would be highly specific for a given amphiphile and was consequently not considered.
The parameters $d_\text{H}$ and $d_\text{BB}$ exhibit broad probability distributions with no well-defined maximum. In turn $C_0$ has a peaked probability distribution yielding a well-defined estimate value and uncertainty.

Here, we discuss for illustration purposes the correlation between $C_0$ and $d_\text{H}$ (Fig.~\ref{fig:corr}a). 
Solutions  along the diagonal line give similar scattering intensities, but lead to significantly different electron density profiles, see Fig.~\ref{fig:corr}b,c. 
%
\begin{figure}\label{fig:corr}
    \includegraphics[width=\textwidth]{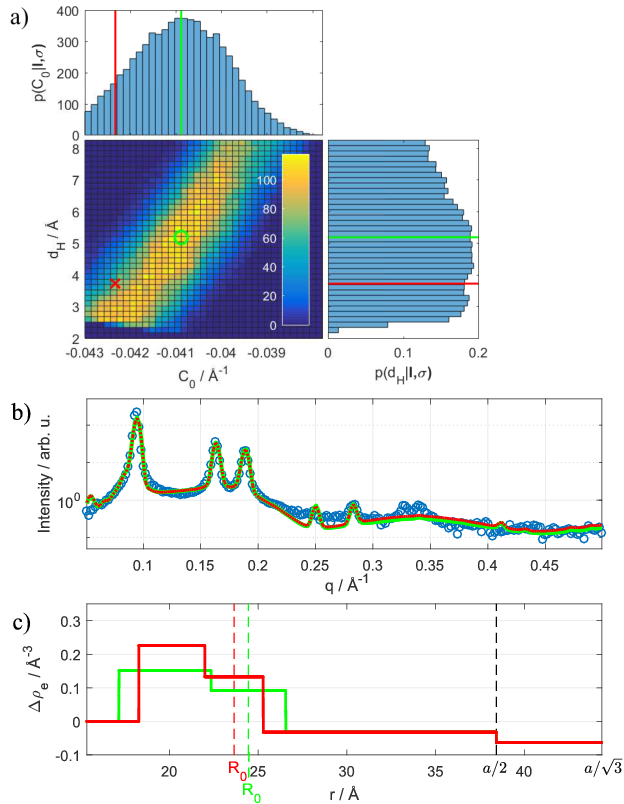}
\caption{Marginal posterior distributions $p(C_0|\mathbf{I},\bm{{\sigma}},\BK)$ and $p(d_{\text{H}}|\mathbf{I},\bm{{\sigma}},\BK)$ of intrinsic curvature $C_0$ and headgroup width $d_{{\text{H}}}$ and $p(C_0,d_\text{H}|\mathbf{I},\bm{{\sigma}},\BK)$ (Panel a). The red cross and corresponding lines mark the sample with the lowest $\chi^2$ (MAP solution), the green circle shows the mean value of the distribution. Panel b) and c) show the corresponding fits and SLD profiles.}
\end{figure}
The correlation between $d_\text{H}$ and $C_0$ may appear counterintuitive. From geometric/physical arguments follows that small $d_\text{H}$ values represent a bending of the phosphate-ethanolamine director of lipid headgroup toward the polar/apolar interface, which leads to a shift of $C_0$ toward more positive values. However, this would lead to significantly different scattered intensities and hence to non-optimal solutions. The mathematical algorithm therefore aims to compensate for this by decreasing $C_0$ for small $d_\text{H}$.
The abrupt drop of the headgroup thickness probability distribution at small $d_\text{H}$ is due to the termination criterion (Eq.~(\ref{eq:nw})).


In the following, we discuss some expectation values $\langle\bfx\rangle$ and the errors $\sigma_\bfx$ obtained by applying Eqs.~(\ref{eq:mean}) and (\ref{eq:error_estimate}). Table \ref{tab:comp_DOPE} compares the obtained structural parameters of DOPE to existing literature values. Our results are in good agreement with previous reports, given the different additives (alkanes or alkenes, some did not use any filler molecule) and slight variations in temperatures. Note that in some cases $A_0$ and $C_0$ have been reported for the pivotal plane. The pivotal plane marks the position within the lipids where the molecular area does not change upon deformation and is usually slightly closer to hydrocarbon tails than the neutral plane~\cite{Leikin.1996,Kollmitzer.2013}. This leads to slight shift of $C_0$ toward positive values.

\begin{table}
\label{tab:comp_DOPE}
\caption{Comparison of structural parameters of DOPE to literature values.}
\begin{tabular}{cccccc|l}
	 $a$ /\AA & $V_\text{L}$ /\AA$^3$  & $d_\text{HH}$ /\AA & $R_\text{W}$ /\AA & $A_{0}$ /\AA$_0^{2}$ & $C_{0}$ /\AA$^{-1}$ &reference \\
    \hline
    76.9 $\pm$ 0.2 & 1142 $\pm$ 10 & 32.4 $\pm$ 1.3 & 22.2 $\pm$ 0.7 & 62.2 $\pm$ 6.0 & -0.0409 $\pm$ 0.0010 & this work\\
    71.9 & 1224 & 31.8 & 20.0 & - & - & \cite{Tate.1989}$^a$\\
	74.9 & -  & - & 22 & - &  -0.033$^i$  & \cite{Rand.1990}$^b$\\
    72.75 & -  & - & 19.1 & - &  - & \cite{Turner.1992}$^c$ \\
     - & - & - & - & - & -0.0367 $\pm$ 0.0005 & \cite{Leikin.1996}$^d$\\
     72.9 & 1220 & 36.0 & 20.4 & 47.4$^i$ & - & \cite{Harper.2001}$^e$\\
     76 & - & - & - &  51.5$^i$& -0.031$^i$ & \cite{DiGregorio.2005}$^f$\\
     - & - & - & - & - & -0.0399 $\pm$ 0.0005 & \cite{Kollmitzer.2013}$^g$\\
     - & - & - & - & - & -0.0365 $\pm$ 0.0012& \cite{Chen.2015}$^h$\\
    \hline
\end{tabular} \\
$^a$ $T = 35^\circ$C, dodecane \\
$^b$ $T = 22^\circ$C, tetradecane \\
$^c$ $T = 30^\circ$C, dodecane \\
$^d$ $T = 25^\circ$C \\
$^e$ $T = 30^\circ$C \\
$^f$ $T = 25^\circ$C \\
$^g$ $T = 35^\circ$C, tricosene \\
$^h$ $T = 30^\circ$C, tetradecane \\
$^i$ determined at the pivotal plane\\
\end{table}

The most direct comparison of $C_0$ can be made to our previous work~\cite{Kollmitzer.2013}, which was performed at the same temperature and tricosene content. Here, we find that the global model combined with  Bayesian analysis yields an intrinsic curvature, which agrees within experimental uncertainty well with our previous result.








\subsection{Effect of temperature}

\begin{figure}
\label{fig:dope_results}
\includegraphics[width=\textwidth]{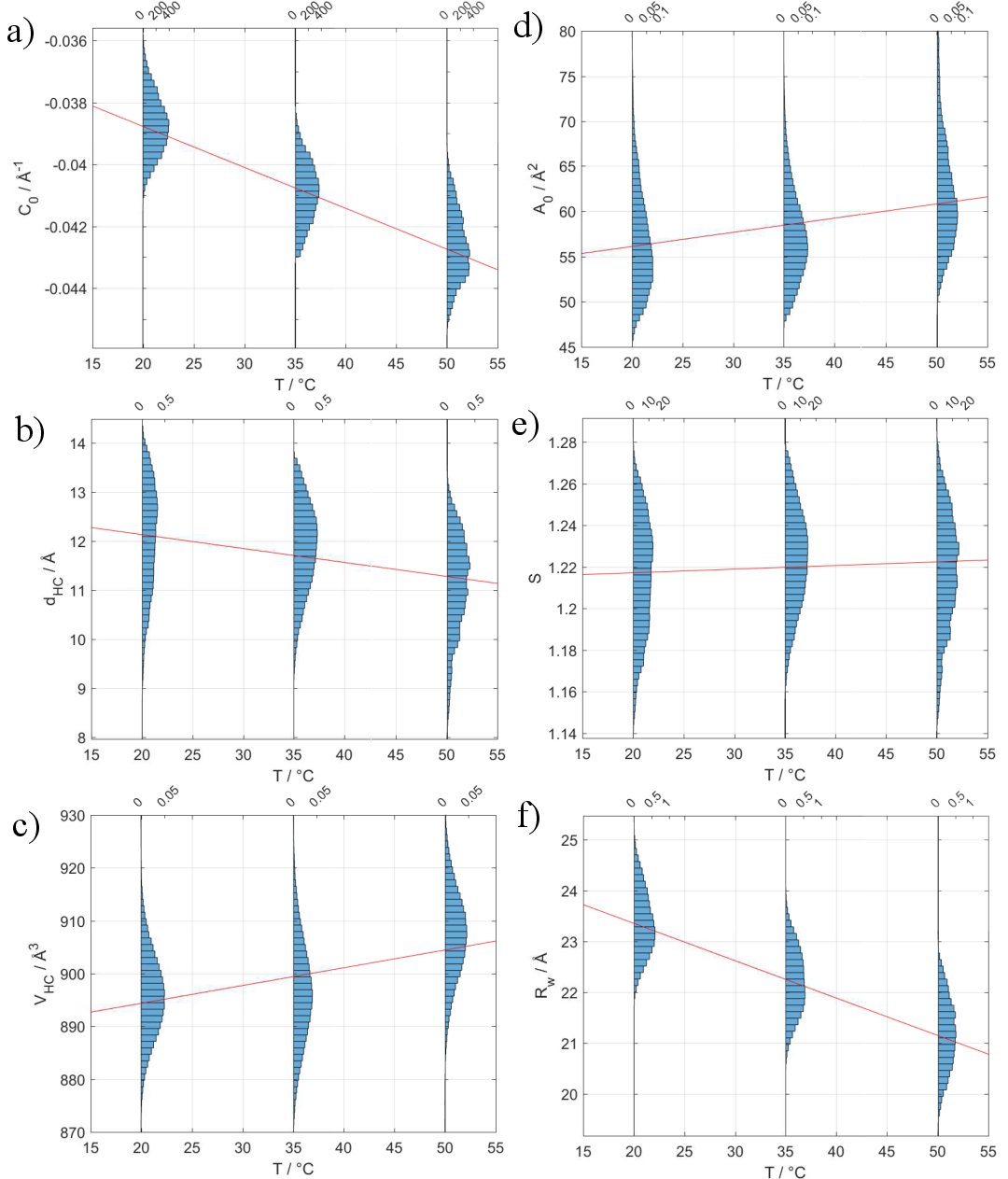}
\caption{Structural parameters of DOPE H$_\text{II}$ as a function of temperature resulting from the Bayesian analysis. Panel a) shows probability densities of the intrinsic curvatures, b) of the hydrocarbon chain length c) of the hydrocarbon chain volume d) of the area per lipid at the neutral plane e) the shape parameter and f) the radius of the water cylinder. Red lines indicate linear regressions of the probability density distributions.}
\end{figure}

Increasing temperature for DOPE should yield a decrease of lipid chain length and concomitant significant increase of the area per lipid at the methyl terminus leading to more negative intrinsic curvatures as reported previously~\cite{Turner.1992,Harper.2001,Tate.1989,Kollmitzer.2013}.
Indeed, our analysis yielded a linear decrease of $C_0$ and $d_\text{HC}$ (Fig. \ref{fig:dope_results}). The slope $\Delta C_0$/$\Delta T = (-1.323 \pm 0.001) \times 10^{-4}$ (\AA~K)$^{-1}$ is identical to our previously reported value~\cite{Kollmitzer.2013}. The relative change of the chain length is in turn $\Delta d_\text{HC}$/$\Delta T = (- 0.0188 \pm 0.0001)$ \AA/K. Interestingly, the shape parameter $\hat{S}$ shows only a modest increase of $\Delta \hat{S}$/$\Delta T = (1.75 \pm 0.03)\times 10^{-4}$ K$^{-1}$, despite the more negative $C_0$ values at higher temperatures and despite the decrease of $d_\text{HC}$ and increase of $V_\text{HC}$ ($\Delta V_\text{HC}$/$\Delta T = (0.3364 \pm 0.0010)$ \AA$^3$/K). This results from a concomitant increase of headgroup area ($\Delta A_0$/$\Delta T = (0.1567 \pm 0.0006)$ \AA$^2$/K) at the neutral plane -- and analogously also at the position of the polar/apolar interface $(R_0 + \frac{d_\text{BB}}{2})$ --, which compensates for the changes of $d_\text{HC}$ and $V_\text{HC}$. The radius of the water core decreases with ($\Delta R_\text{W}$/$\Delta T = (-7.364 \pm 0.007)\times 10^{-2}$ \AA/K).


\subsection{Effect of hydrocarbon chain composition}
Finally, we tested the applicability of the analysis technique to PEs with differing hydrocarbon chain composition.  In particular, we studied the H$_\text{II}$ phases of POPE, which has a palmitoyl and an oleoyl chain, DMPE, which has two myristoyl chains, and diC16:1PE, with two palmitoeloyl hydrocarbons. Note that pure POPE forms a H$_\text{II}$ phase only above 71$^\circ$C, while the lamellar to H$_\text{II}$ phase transition temperature $T_H$ for pure di16:1PE was reported to be $43.4^\circ$C and $T_H> 100^\circ$C for DMPE \cite{Koynova.1994}. The addition of alkanes or alkenes to inverted hexagonal phases is known to reduce stress resulting from interstitial space between the individual rods~\cite{Vacklin.2000,Chen.1998,Kirk.1985}. We  previously demonstrated that tricosene sufficiently lowers the $T_\text{H}$ for POPE to perform a H$_\text{II}$ phase analysis at physiological temperature~\cite{Kollmitzer.2013}. Similarly, di16:1PE formed a neat H$_\text{II}$ phase at 35$^\circ$C upon adding 12 wt\% tricosene (see below). In the case of DMPE we found a pure H$_\text{II}$ scattering pattern only for $T \ge 80^\circ$C, indicating a significantly less negative $C_0$. For this reason we performed the global analysis at 35$^\circ$C for POPE and di16:1PE and at 80$^\circ$C for DMPE.

\begin{figure}\label{fig:fits}
\caption{Data points (including error bars) and expectation value of the intensity (with error bands) of SAXS patterns of POPE (35$^\circ$C), di16:1PE (35$^\circ$C) and DMPE (80$^\circ$C). 
}
\includegraphics[width=\textwidth]{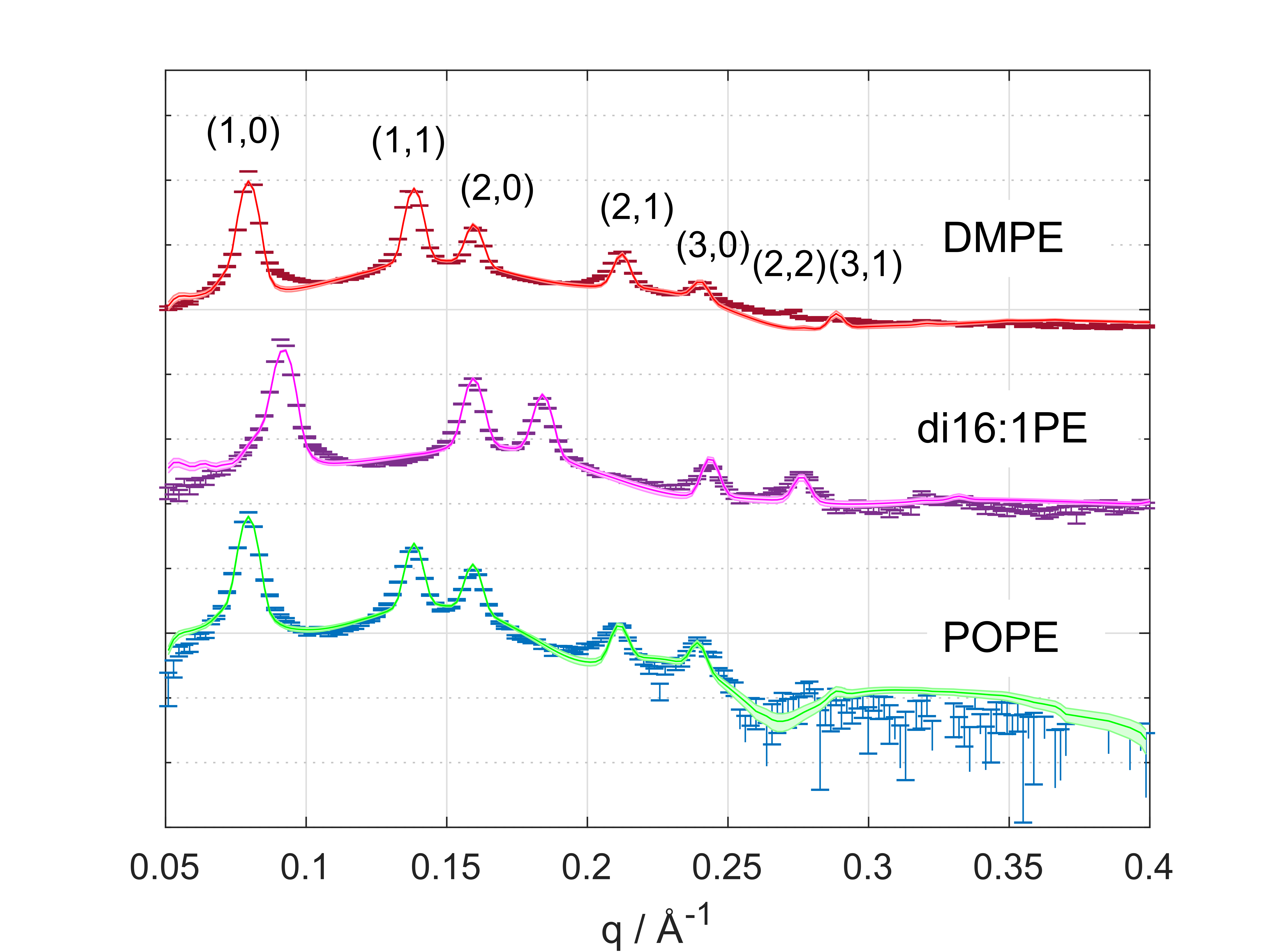}
\end{figure}

\begin{figure}
\label{fig:all_results}
		\caption{Intrinsic curvature (a) and hydrocarbon chain length (b) probability densities and mean values (red) for various lipids at 35 $^\circ$C (except DMPE: 80 $^\circ$C).}
		\includegraphics[width=0.7\textwidth]{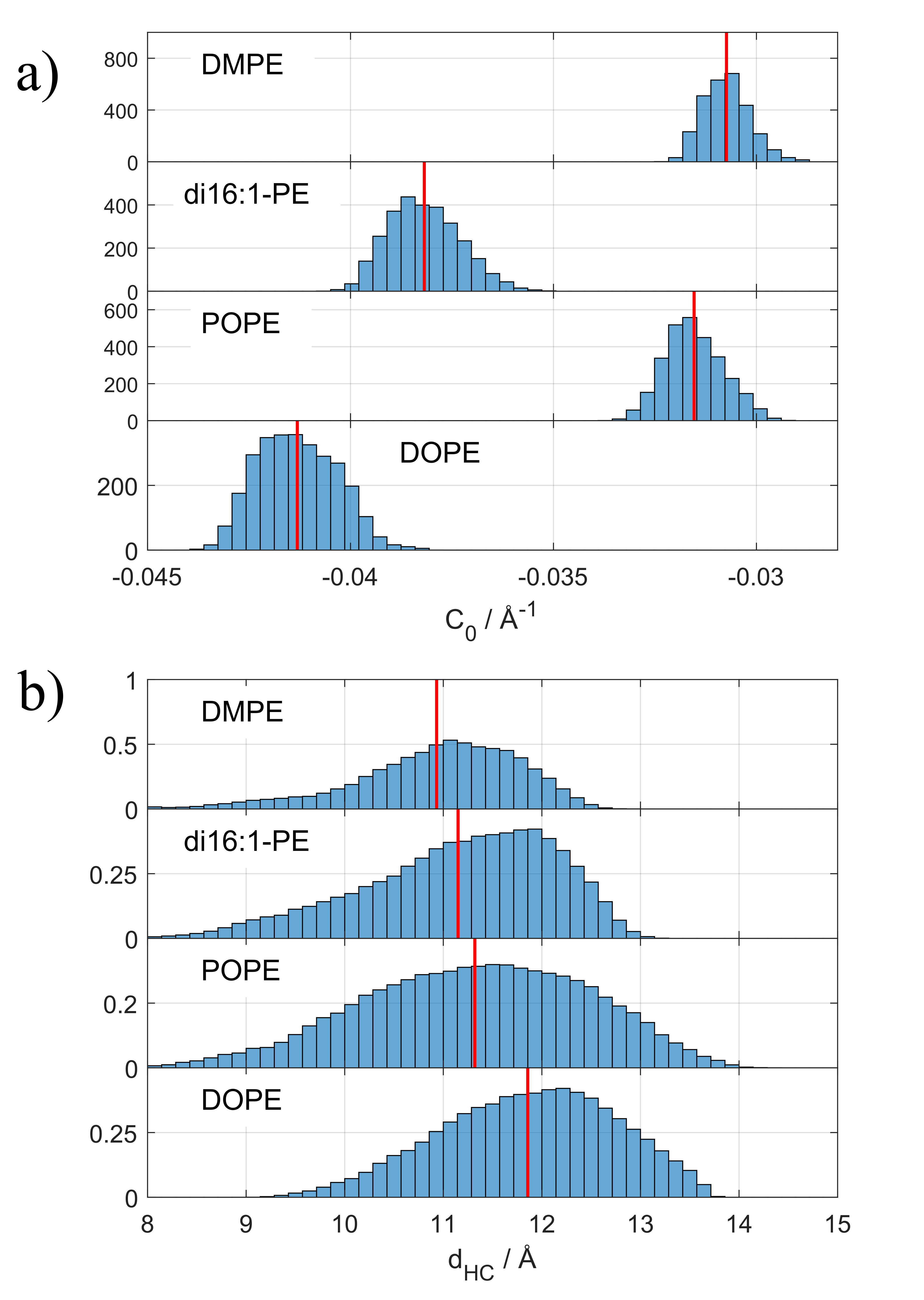}
\end{figure}

Unlike DOPE, the (2,1)-peak was clearly present in the scattering data of all three lipids, a feature which helped to obtain a better agreement of the model with experimental data (Fig.~\ref{fig:fits}). The corresponding probability density distributions for $C_0$ clearly show that monounsaturated hydrocarbons induce significantly more negative intrinsic curvature than saturated hydrocarbon, which is due to the kink induced by the \textit{cis}-double bond. The proximity of values for POPE and DMPE is attributed to the temperature difference and the associated decrease of $C_0$ (Fig.~\ref{fig:all_results}). Assuming a similar temperature dependence as observed for DOPE yields a $C_0$ close to zero  for DMPE at 35$^\circ$C, which agrees with the well-established observation that DMPE prefers to form bilayers at ambient temperatures. Beside the difference between saturated and unsaturated hydrocarbons our analysis also clearly shows that $C_0^\text{DOPE} < C_0^\text{diC16:1PE}$. That is, increasing the chain length of monounsaturated acyl chains also leads to a more negative $C_0$ value. This signifies that the kink induced by the \textit{cis}-double bond leads to a progressive increase of hydrocarbon splay  upon acyl chain extension.

The mean values of hydrocarbon chain length show a trend in the expected direction, i.e. they increase with the number of hydrocarbons (Fig. \ref{fig:all_results}), but all cases exhibit a broad distribution as a result of the not well-defined backbone width $d_\text{BB}$.





The expectation values for $C_0$ and $d_\text{HC}$ for the different lipids are listed in Tab.~\ref{tab:res}, including resulting structural parameters for $R_\text{W}$, $A_0$, $V_\text{HC}$, and $\tilde{S}$. Previously, we reported $C_0 = -0.0316$ \AA$^{-1}$ for POPE at 37$^\circ$C~\cite{Kollmitzer.2013}, which is in excellent agreement with our present value. Regarding other structural parameters we particularly found that $R_\text{W}$ and $A_0$ decrease with $C_0$ becoming more negative, which is mainly attributed to the geometry of the H$_\text{II}$ phase. 
The hydrocarbon chain volumes are in agreement with the chemical compositions. That is, DMPE with two C14:0 chains has the smallest and DOPE with two C18:1 chains has the largest $V_\text{HC}$ value, whereas volumes of POPE and diC16:1PE take up intermediate values. Our hydrocarbon volume of POPE is about 4 \% lower than the value reported for POPE in the absence of tricosene at the same temperature, where it forms a fluid lamellar phase \cite{Kucerka.2015}. This indicates a slightly tighter hydrocarbon chain packing in fully relaxed monolayers.
The shape parameter (DMPE $\simeq$ POPE $<$ diC16:1PE $<$ DOPE) clearly shows that from all lipids presently studied DOPE has the highest propensity to form a H$_\text{II}$ phase, which is consistent with its low $T_\text{H}$~\cite{Koynova.1994}.

\begin{table}\label{tab:res}
\caption{Comparison of structural parameters of different phosphatidylethanolamines.}
\begin{tabular}{l|cccccc}      
	& $C_0$ / \AA$^{-1}$&	$d_\text{HC}$ / \AA & $R_\text{W}$ / \AA &$A_0$ / \AA$^2$ & $V_\text{HC}$ / \AA$^3$ & $\tilde{S}$ \\
    \hline
DMPE$^a$&	$-0.0314\pm 0.0006$ &	$10.9 \pm 0.9$ & $30.6 \pm 0.7$ &	$59 \pm 6$ & $737 \pm 7$ & $1.16 \pm 0.02$\\
di16:1PE$^b$&	$-0.0382 \pm 0.0009$  &	$11.1 \pm 1.0$ & $24.0 \pm 0.7$ &	$60 \pm 7$ & $797 \pm 8$ & $1.20\pm 0.03$\\
POPE$^b$&	$-0.0317 \pm 0.0007$ &	$11.3 \pm 1.2$& $29.2 \pm 0.9$ &	$68 \pm 9$ & $884 \pm 9$ & $1.17 \pm 0.03$\\
DOPE$^b$ &$-0.0409 \pm 0.0010$&	$11.9 \pm 0.9$ & $22.2 \pm 0.7$ &	$62 \pm 6$ & $897 \pm 10$ & $1.22 \pm 0.03$\\
\hline
\end{tabular} \\
$^a$ $T = 80^\circ$C \\
$^b$ $T = 35^\circ$C
\end{table}
     
\section{Conclusions}
We have introduced a global scattering model for fully hydrated unoriented H$_\text{II}$ phases. Compared to previous models for H\textsubscript{I} phases~\cite{Freiberger.2006,Sundblom.2009}, H$_\text{II}$ phase analysis required to add diffuse scattering not originating from hexagonal structures. While the exact origin of this additional contribution remains unclear, we successfully modeled the measured SAXS pattern upon including a lamellar form factor. The SLD of the lipid unit cell was constrained by compositional modeling using complementary information on lipid volume and structure. This description is generic and entails the analysis of SAXS and small-angle neutron scattering (SANS) data. In particular a joint analysis of SAXS and differently contrasted SANS data (see, e.g.~\cite{Pabst.2010,Heberle.2017}) might be beneficial for increased structural resolution regarding the lipid head and backbone groups.

Here, we analyzed SAXS data using Bayesian probability theory combined with MCMC simulations. This was specifically necessary due to the weakly-defined global minimum of the optimization cost function.
The full probabilistic approach provides the probability density distributions of the involved parameters leading to reliable parameter estimates including errors.

The obtained estimates are in good agreement with previously reported structural data of DOPE and POPE. We further provided details for lipid structures of DMPE and di16:1PE in the H$_\text{II}$ phase, clearly demonstrating that out of all presently studied lipids DMPE is least prone to form a H$_\text{II}$ phase. The developed technique will be easily transferred to other H$_\text{II}$ phase amphiphiles using appropriate compositional modeling. In particular, we are envisioning a high potential for applications in drug-delivery formulations involving H$_\text{II}$ structures, which exhibit only weak Bragg peaks, but significant contributions from diffuse scattering, such as hexosomes~(see, e.g. \cite{Yaghmur.2009}). Another potential application is the determination of intrinsic lipid curvatures of lamellar-phase-forming lipids using mixtures with DOPE~\cite{Kollmitzer.2013}, which is particularly encouraged by the high robustness of the retrieved $C_0$ estimates. Such approaches are currently being explored in our laboratory.

\section{Acknowledgements}
We thank D. Kopp, M. Pachler and J. Kremser for technical assistance, and E. Semeraro for critical reading of the manuscript. We further thank O. Glatter for performing a trial analysis using his GIFT software package. This work was supported financially by the Austrian Science Funds FWF (grant no. P27083-B20 to G.P.)





\bibliographystyle{ieeetr}
\bibliography{globhex}

\begin{thebibliography}{10}

\bibitem{Glatter.2018}
O.~Glatter, {\em Scattering methods and their application in colloid and
  interface science}.
\newblock Amsterdam: Elsevier, 2018.

\bibitem{Heberle.2017}
F.~A. Heberle and G.~Pabst, ``Complex biomembrane mimetics on the sub-nanometer
  scale,'' {\em Biophys Rev}, vol.~9, no.~4, pp.~353--373, 2017.

\bibitem{Koltover.1998}
I.~Koltover, ``An inverted hexagonal phase of cationic liposome-dna complexes
  related to dna release and delivery,'' {\em Science}, vol.~281, no.~5373,
  pp.~78--81, 1998.

\bibitem{Yaghmur.2009}
A.~Yaghmur and O.~Glatter, ``Characterization and potential applications of
  nanostructured aqueous dispersions,'' {\em Adv Colloid Interface Sci},
  vol.~147-148, pp.~333--342, 2009.

\bibitem{Leikin.1996}
S.~Leikin, M.~M. Kozlov, N.~L. Fuller, and R.~P. Rand, ``Measured effects of
  diacylglycerol on structural and elastic properties of phospholipid
  membranes,'' {\em Biophys J}, vol.~71, no.~5, pp.~2623--2632, 1996.

\bibitem{DiGregorio.2005}
G.~M. {Di Gregorio} and P.~Mariani, ``Rigidity and spontaneous curvature of
  lipidic monolayers in the presence of trehalose: A measurement in the dope
  inverted hexagonal phase,'' {\em Eur Biophys J}, vol.~34, no.~1, pp.~67--81,
  2005.

\bibitem{Kollmitzer.2013}
B.~Kollmitzer, P.~Heftberger, M.~Rappolt, and G.~Pabst, ``Monolayer spontaneous
  curvature of raft-forming membrane lipids,'' {\em Soft Matter}, vol.~9,
  no.~45, pp.~10877--10884, 2013.

\bibitem{Chen.2015}
Y.-F. Chen, K.-Y. Tsang, W.-F. Chang, and Z.-A. Fan, ``Differential
  dependencies on ca2+ and temperature of the monolayer spontaneous curvatures
  of dope, dopa and cardiolipin: Effects of modulating the strength of the
  inter-headgroup repulsion,'' {\em Soft Matter}, vol.~11, no.~20,
  pp.~4041--4053, 2015.

\bibitem{Kozlov.1991}
M.~M. Kozlov and M.~Winterhalter, ``Elastic moduli for strongly curved
  monoplayers. position of the neutral surface,'' {\em J Phys II}, vol.~1,
  no.~9, pp.~1077--1084, 1991.

\bibitem{Marsh.2006}
D.~Marsh, ``Elastic curvature constants of lipid monolayers and bilayers,''
  {\em Chem Phys Lipids}, vol.~144, no.~2, pp.~146--159, 2006.

\bibitem{Dan.1998}
N.~Dan and S.~A. Safran, ``Effect of lipid characteristics on the structure of
  transmembrane proteins,'' {\em Biophys J}, vol.~75, no.~3, pp.~1410--1414,
  1998.

\bibitem{Frewein.2016}
M.~Frewein, B.~Kollmitzer, P.~Heftberger, and G.~Pabst, ``Lateral
  pressure-mediated protein partitioning into liquid-ordered/liquid-disordered
  domains,'' {\em Soft Matter}, vol.~12, no.~13, pp.~3189--3195, 2016.

\bibitem{Frolov.2011}
V.~A. Frolov, A.~V. Shnyrova, and J.~Zimmerberg, ``Lipid polymorphisms and
  membrane shape,'' {\em Cold Spring Harb Perspect Biol}, vol.~3, no.~11,
  p.~a004747, 2011.

\bibitem{Tate.1989}
M.~W. Tate and S.~M. Gruner, ``Temperature dependence of the structural
  dimensions of the inverted hexagonal (hii) phase of
  phosphatidylethanolamine-containing membranes,'' {\em Biochemistry}, vol.~28,
  no.~10, pp.~4245--4253, 1989.

\bibitem{Turner.1992}
D.~C. Turner and S.~M. Gruner, ``X-ray diffraction reconstruction of the
  inverted hexagonal (hii) phase in lipid-water systems,'' {\em Biochemistry},
  vol.~31, no.~5, pp.~1340--1355, 1992.

\bibitem{Rand.1990}
R.~P. Rand, N.~L. Fuller, S.~M. Gruner, and V.~A. Parsegian, ``Membrane
  curvature, lipid segregation, and structural transitions for phospholipids
  under dual-solvent stress,'' {\em Biochemistry}, vol.~29, no.~1, pp.~76--87,
  1990.

\bibitem{Harper.2001}
P.~E. Harper, D.~A. Mannock, R.~N. Lewis, R.~N. McElhaney, and S.~M. Gruner,
  ``X-ray diffraction structures of some phosphatidylethanolamine lamellar and
  inverted hexagonal phases*,'' {\em Biophys J}, vol.~81, no.~5,
  pp.~2693--2706, 2001.

\bibitem{Pabst.2000}
G.~Pabst, M.~Rappolt, H.~Amenitsch, and P.~Laggner, ``Structural information
  from multilamellar liposomes at full hydration: Full q -range fitting with
  high quality x-ray data,'' {\em Phys Rev E}, vol.~62, no.~3, pp.~4000--4009,
  2000.

\bibitem{Freiberger.2006}
N.~Freiberger and O.~Glatter, ``Small-angle scattering from hexagonal liquid
  crystals,'' {\em J Phys Chem B}, vol.~110, no.~30, pp.~14719--14727, 2006.

\bibitem{Sundblom.2009}
A.~Sundblom, C.~L.~P. Oliveira, A.~E.~C. Palmqvist, and J.~S. Pedersen,
  ``Modeling in situ small-angle x-ray scattering measurements following the
  formation of mesostructured silica,'' {\em J Phys Chem C}, vol.~113, no.~18,
  pp.~7706--7713, 2009.

\bibitem{Buboltz.1999}
J.~T. Buboltz and G.~W. Feigenson, ``A novel strategy for the preparation of
  liposomes: Rapid solvent exchange,'' {\em Biochim Biophys Acta, Biomembr},
  vol.~1417, no.~2, pp.~232--245, 1999.

\bibitem{Leber.2018}
R.~Leber, M.~Pachler, I.~Kabelka, I.~Svoboda, D.~Enkoller, R.~V{\'a}cha,
  K.~Lohner, and G.~Pabst, ``Synergism of antimicrobial frog peptides couples
  to membrane intrinsic curvature strain,'' {\em Biophys J}, vol.~114, no.~8,
  pp.~1945--1954, 2018.

\bibitem{Rieder.2015}
A.~A. Rieder, D.~Koller, K.~Lohner, and G.~Pabst, ``Optimizing rapid solvent
  exchange preparation of multilamellar vesicles,'' {\em Chem Phys Lipids},
  vol.~186, pp.~39--44, 2015.

\bibitem{Alley.2008}
S.~H. Alley, O.~Ces, M.~Barahona, and R.~H. Templer, ``X-ray diffraction
  measurement of the monolayer spontaneous curvature of
  dioleoylphosphatidylglycerol,'' {\em Chem Phys Lipids}, vol.~154, no.~1,
  pp.~64--67, 2008.

\bibitem{Oster.1952}
G.~Oster and D.~P. Riley, ``Scattering from cylindrically symmetric systems,''
  {\em Acta Cryst}, vol.~5, no.~2, pp.~272--276, 1952.

\bibitem{Marchal.2003}
D.~Marchal and B.~Dem{\'e}, ``Small-angle neutron scattering by porous alumina
  membranes made of aligned cylindrical channels,'' {\em J Appl Crystallogr},
  vol.~36, no.~3, pp.~713--717, 2003.

\bibitem{Forster.2005}
S.~F{\"o}rster, A.~Timmann, M.~Konrad, C.~Schellbach, A.~Meyer, S.~S. Funari,
  P.~Mulvaney, and R.~Knott, ``Scattering curves of ordered mesoscopic
  materials,'' {\em J Phys Chem B}, vol.~109, no.~4, pp.~1347--1360, 2005.

\bibitem{Szekely.2010}
P.~Sz{\'e}kely, A.~Ginsburg, T.~Ben-Nun, and U.~Raviv, ``Solution x-ray
  scattering form factors of supramolecular self-assembled structures,'' {\em
  Langmuir}, vol.~26, no.~16, pp.~13110--13129, 2010.

\bibitem{Franks.1982}
N.~P. Franks, V.~Melchior, D.~A. Kirschner, and D.~Caspar, ``Structure of
  myelin lipid bilayers,'' {\em J Mol Biol}, vol.~155, no.~2, pp.~133--153,
  1982.

\bibitem{Ding.2004}
L.~Ding, W.~Liu, W.~Wang, C.~J. Glinka, D.~L. Worcester, L.~Yang, and H.~W.
  Huang, ``Diffraction techniques for nonlamellar phases of phospholipids,''
  {\em Langmuir}, vol.~20, no.~21, pp.~9262--9269, 2004.

\bibitem{Kucerka.2015}
N.~Ku{\v{c}}erka, B.~{van Oosten}, J.~Pan, F.~A. Heberle, T.~A. Harroun, and
  J.~Katsaras, ``Molecular structures of fluid phosphatidylethanolamine
  bilayers obtained from simulation-to-experiment comparisons and experimental
  scattering density profiles,'' {\em J Phys Chem B}, vol.~119, no.~5,
  pp.~1947--1956, 2015.

\bibitem{Israelachvili.2011}
J.~N. Israelachvili, {\em Intermolecular and Surface Forces}.
\newblock {Academic Press}, 2011.

\bibitem{Jaynes.2003}
E.~T. Jaynes and G.~L. Bretthorst, {\em Probability theory: The logic of
  science}.
\newblock 2003.

\bibitem{Gregory.2005}
P.~Gregory, {\em Bayesian Logical Data Analysis for the Physical Sciences}.
\newblock Cambridge: Cambridge University Press, 2005.

\bibitem{Sivia.2012}
D.~S. Sivia and J.~H. Skillings, {\em Data analysis: A Bayesian tutorial ; [for
  scientists and engineers]}.
\newblock Oxford science publications, Oxford: {Oxford Univ. Press}, second
  edition~ed., 2012.

\bibitem{Linden.2014}
W.~von~der Linden, V.~Dose, and U.~von Toussaint, {\em Bayesian Probability
  Theory}.
\newblock Cambridge: {Cambridge University Press}, 2014.

\bibitem{Jeffreys.1946}
H.~Jeffreys, ``An invariant form for the prior probability in estimation
  problems,'' {\em Proc R Soc London, Ser A}, vol.~186, no.~1007, pp.~453--461,
  1946.

\bibitem{Koynova.1994}
R.~Koynova and M.~Caffrey, ``Phases and phase transitions of the hydrated
  phosphatidylethanolamines,'' {\em Chem Phys Lipids}, vol.~69, no.~1,
  pp.~1--34, 1994.

\bibitem{Vacklin.2000}
H.~Vacklin, B.~J. Khoo, K.~H. Madan, J.~M. Seddon, and R.~H. Templer, ``The
  bending elasticity of 1-monoolein upon relief of packing stress,'' {\em
  Langmuir}, vol.~16, no.~10, pp.~4741--4748, 2000.

\bibitem{Chen.1998}
Z.~Chen and R.~P. Rand, ``Comparative study of the effects of several n-alkanes
  on phospholipid hexagonal phases,'' {\em Biophys J}, vol.~74, no.~2,
  pp.~944--952, 1998.

\bibitem{Kirk.1985}
G.~L. Kirk and S.~M. Gruner, ``Lyotropic effects of alkanes and headgroup
  composition on the la -hii lipid liquid crystal phase transition:
  Hydrocarbon packing versus intrinsic curvature,'' {\em J Phys}, vol.~46,
  no.~5, pp.~761--769, 1985.

\bibitem{Pabst.2010}
G.~Pabst, N.~Kucerka, M.-P. Nieh, M.~C. Rheinst{\"a}dter, and J.~Katsaras,
  ``Applications of neutron and x-ray scattering to the study of biologically
  relevant model membranes,'' {\em Chem Phys Lipids}, vol.~163, no.~6,
  pp.~460--479, 2010.

\end{thebibliography}




\end{document}